\documentclass[preprint2]{aastex}

\usepackage{hyperref}
\usepackage{cleveref}
\usepackage{epsfig,psfig}
\usepackage{amsmath}
\usepackage{array,longtable}

\newcommand{\E}[1]{\times 10^{#1}}
      \newcommand{\ps}{\,{\rm s}$^{-1}$}
\newcommand{\yr}{\,{\rm yr}}    \newcommand{\Msun}{{\rm M}_{\rm \odot}}
\newcommand{\cm}{\,{\rm cm}}    \newcommand{\km}{\,{\rm km}}

\newcommand{\ncol}{$N({\rm H_2})$} \newcommand{\tex}{$T_{\rm ex}$}
\newcommand{\brightnessunit}{W~m$^{-2}$~Hz$^{-1}$~sr$^{-1}$}
\newcommand{\degree}{^{\circ}}

\newcommand{\du}{d_{3.1}} 
\newcommand{\dua}{d_{0.31}} 
\newcommand{\HI}{H~{\sc i}}
\newcommand{\vlsr}{$V_{\rm LSR}$}       \newcommand{\tmb}{$T_{\rm mb}$}
\newcommand{\twCO}{$^{12}$CO}  \newcommand{\thCO}{$^{13}$CO}
\newcommand{\CeiO}{C$^{18}$O} 
\newcommand{\snr}{{G43.9+1.6}}

\shorttitle{SNR G43.9+1.6}
\shortauthors{Zhou et al.}

\begin{document}

\title{Molecular Clouds Surrounding Supernova Remnant G43.9+1.6: Associated and Non-associated}

\author{Xin Zhou\altaffilmark{1,2}, Yang Su\altaffilmark{1,2}, Ji Yang\altaffilmark{1,2}, Yang Chen\altaffilmark{3,4}, Ye Xu\altaffilmark{1,2}, 
Xuepeng Chen\altaffilmark{1,2},
Shaobo Zhang\altaffilmark{1,2} 
}
\affil{$^1$Purple Mountain Observatory, Chinese Academy of Sciences, No.10 Yuanhua Road, Nanjing 210033, China; xinzhou@pmo.ac.cn \\
$^2$Key Laboratory of Radio Astronomy, Chinese Academy of Sciences, Nanjing 210033, China \\
$^3$Department of Astronomy, Nanjing University, 163 Xianlin Avenue, Nanjing 210023, China \\
$^4$Key Laboratory of Modern Astronomy and Astrophysics, Nanjing University, Ministry of Education, Nanjing 210023, China}

\email{xinzhou@pmo.ac.cn}

\begin{abstract}
Many supernova remnants (SNRs) are considered to evolve in molecular environments, but the associations between SNRs and molecular clouds (MCs) are often unclear.
Being aware of such ambiguous case, we report our study on the molecular environment towards the SNR~\snr\ by CO line observations.
We investigated the correlations between the SNR and MCs at different velocities, and found two velocity components, i.e.\ $\sim$5~\km\ps\ and $\sim$50~\km\ps\ velocity components, showing spatial correlations with the remnant.
However, no dynamical evidence of disturbance was found for the $\sim$5~\km\ps\ velocity component.
At the distance of the $\sim$5~\km\ps\ velocity component, either near or far distance, the derived physical parameters are unreasonable too.
We conclude that the SNR is not associated with the $\sim$5~\km\ps\ velocity component, and their spatial correlation is just a chance correlation. 
For the $\sim$50~\km\ps\ velocity component, dynamical evidence of disturbances, as well as the spatial correlation, indicate that it is associated with the SNR.
We found that all the CO spectra extracted from the molecular clumps distributed along the border of the remnant are with broadened components presented, which can be fitted by Gaussian functions. 
By further analysis, we suggest that the SNR is at a near kinematic distance of about 3.1~kpc.
\end{abstract}


\section{Introduction}
Remnants of core-collapse supernova are thought to be in the vicinity of molecular clouds (MCs), due to the short lifetimes of their progenitor stars after being born in parent MCs.
Recently observations of Tycho's supernova remnant (SNR) reveal that Type Ia SNR may be also associated with MC \citep{Zhoup+2016, Chen+2017}.
There are about 80 Galactic SNRs which are confirmed or suggested to be in physical contact with MCs, among 294 Galactic SNRs, at present \citep[][and references therein]{
Gaensler+2008, Jiang+2010, Tian+2010, Eger+2011, Jeong+2012, Frail+2013, Chen+2014, Fukuda+2014, Su+2014, Zhou+2014, Zhu+2014, Zhang+2015, Voisin+2016, Zhoup+2016a, Zhou+2016, deWilt+2017, Lau+2017, Liu+2017, Su+2017a, Liu+2018, Maxted+2018, Su+2018, Ma+2019, Yu+2019, Green2019}. 
Based on associations between SNRs and MCs, we could determine the distances of SNRs, and accordingly, the evolutionary stages of SNRs \citep[e.g.][]{Kilpatrick+2016, Su+2017b}.
SNRs associated with MCs are also good laboratories to study the interactions between SNR shocks and molecular gases \citep[e.g.][]{Dickman+1992, Zhou+2014} as well as the acceleration of cosmic rays in SNRs \citep[e.g.][]{AharonianAtoyan1996, Hewitt+2009, LiChen2012}. 
If an MC overlaps with an SNR in position and presents some spatial similarity, it can be considered as a candidate MC in association with the SNR. We may call this case spatial correlation between the MC and the SNR. The possibility of their association would be enhanced if their spatial correlation is strong.
Spatial correlation is commonly used as non-independent evidence for SNR-MC association.
About half of SNR-MC associations are suggested based on their spatial correlations \citep[see Table~2 in][]{Jiang+2010}.
However, such evidence could be problematic, due to overlaps of multiple MCs in different spiral arms in the line-of-sight, especially, toward the inner region of the Milky Way. Molecular gas distributions around SNRs are usually complicated. 
One needs to further examine the SNR-MC association candidates not only on their spatial correlations but also on other more robust evidences, e.g.\ OH~1720~MHz maser emissions, dynamical evidence of interaction, etc. \citep[see discussion in][for reference]{Jiang+2010}.

The SNR~\snr\ is a faint radio source with a diameter of $\sim60'$, which presents a partial shell structure with no obvious plerionic component \citep{Reich+1990, Vasisht+1994}. As indicated by its radio morphology, this SNR can be classified as a possible shell-type SNR, which is supported by its spectral index of $-0.47\pm0.06$ \citep{Reich+1990, Vasisht+1994, Sun+2011}.
There is no direct distance measurement for the SNR.
We present in this work wide-field CO line observations toward the SNR~\snr, in order to investigate the molecular environment of the remnant. Full examinations of spatial distributions as well as further spectral analyses of different velocity components in the line-of-sight are performed.

\section{Observations}\label{sec:obs}
The CO line emission was observed using the Purple Mountain Observatory Delingha (PMODLH) 13.7~m millimeter-wavelength telescope \citep{Zuo+2011}, which is a part of the Milky Way Image Scroll Painting (MWISP)--CO line survey project\footnote{http://www.radioast.nsdc.cn/yhhjindex.php}. 
The \twCO~(J=1--0), \thCO~(J=1--0), and \CeiO~(J=1--0)  lines were observed simultaneously with the $3\times3$ multibeam sideband separation superconducting receiver \citep{Shan+2012}.
We mapped a $2^\circ\times 2^\circ$ area covering the full extent of the SNR~\snr\ via on-the-fly (OTF) observing mode. 
The data were meshed with a grid spacing of $30''$, and the half-power beam width (HPBW) was $\sim51"$.
The spectral resolutions of the three CO lines were 0.17~\km\ps\ for \twCO~(J=1--0) and 0.16~\km\ps\ for both \thCO~(J=1--0) and \CeiO~(J=1--0).
The typical {\it rms} noises were $\sim0.5$~K for \twCO~(J=1--0) and $\sim0.3$~K for \thCO~(J=1--0) and \CeiO~(J=1--0).
All data were reduced using the GILDAS/CLASS package\footnote{http://www.iram.fr/IRAMFR/GILDAS}.
Further observation parameters and data processing details are in \cite{Zhou+2016}.
6~cm radio continuum emission data were also obtained from the Sino-German $\lambda$6~cm polarization survey, of which the angular resolution is $9'.5$ \citep{Sun+2011}.

\section{Results}\label{sec:result}
\begin{figure}[ptbh!]
\centerline{{\hfil\hfil
\psfig{figure=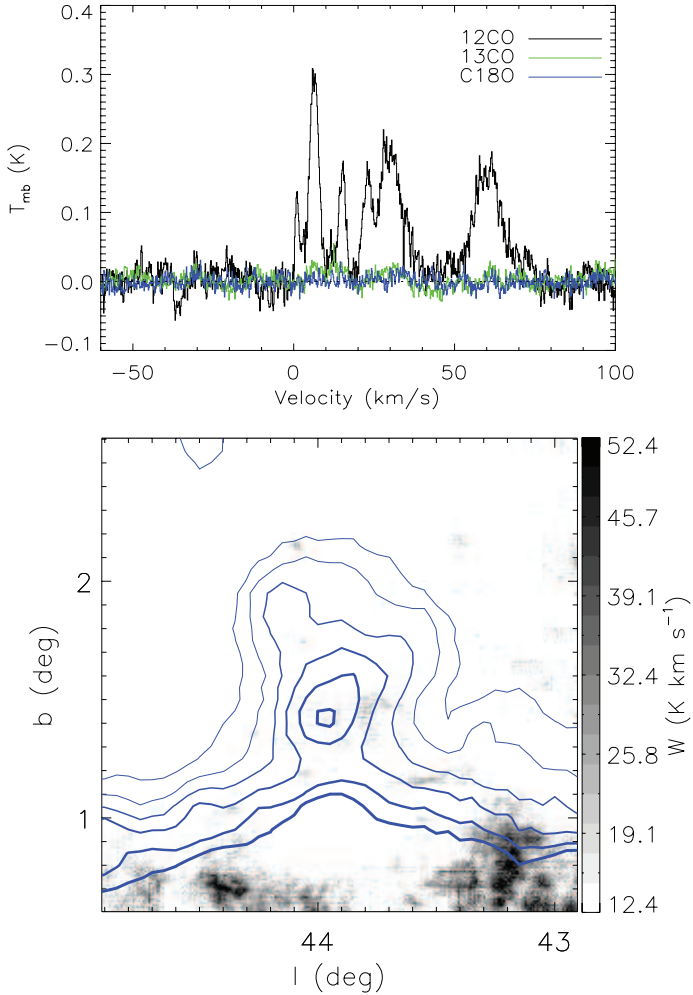 ,height=5.5in,angle=0, clip=}
\hfil\hfil}}
\caption{{\sl upper:} Average CO spectra over the $2^\circ\times 2^\circ$ region. 
{\sl lower:} \twCO~(J=1--0) intensity map integrated over the velocity range of 0--80~\km\ps, overlaid with 6~cm radio continuum emission contours with an angular resolution of $9'.5$ (blue).
The contour levels of the 6~cm radio continuum emission are 50, 66, 82, 98, 114, and 130~mK T$_B$.
The minimum value of \twCO\ emission as shown in the colorbar is $5\sigma$. 
}
\label{f:overall}
\end{figure}

The SNR~\snr\ has partial shell structure in radio continuum emission, of which the ridge and the peak (hereafter radio shell peak) locate on the remnant's near side toward the Galactic plane \citep[see lower panel of Figure~\ref{f:overall};][]{Sun+2011}. Its radio emission in south is contaminated by background radio emission in the Galactic plane, e.g.\ from star-forming regions. In the mapping area of CO observation, we detected several CO velocity components (see {\sl upper} panel of Figure~\ref{f:overall}). The velocity components in the range of 0--20~\km\ps\ are probably from adjacent MCs, i.e.\ belong to the Aquila Rift feature, besides, the other prominent velocity components are mostly from the Sagittarius spiral arm \citep[][and references therein]{Reid+2016}.
There is some CO emission in the remnant region, and more CO emission outside of the remnant region (see lower panel of Figure~\ref{f:overall}). In general, the CO emission is stronger when closer to the Galactic plane.
Note that only prominent CO emissions can be seen in the integrated intensity map over a large velocity range.
We examined all the velocity components, and found no spatial correlation between them and the SNR, except for the components around 5~\km\ps\ and 50\km\ps.

\subsection{The Local Velocity Component}\label{subsec:6vc}
\begin{figure*}[ptbh!]
\centerline{{\hfil\hfil
\psfig{figure=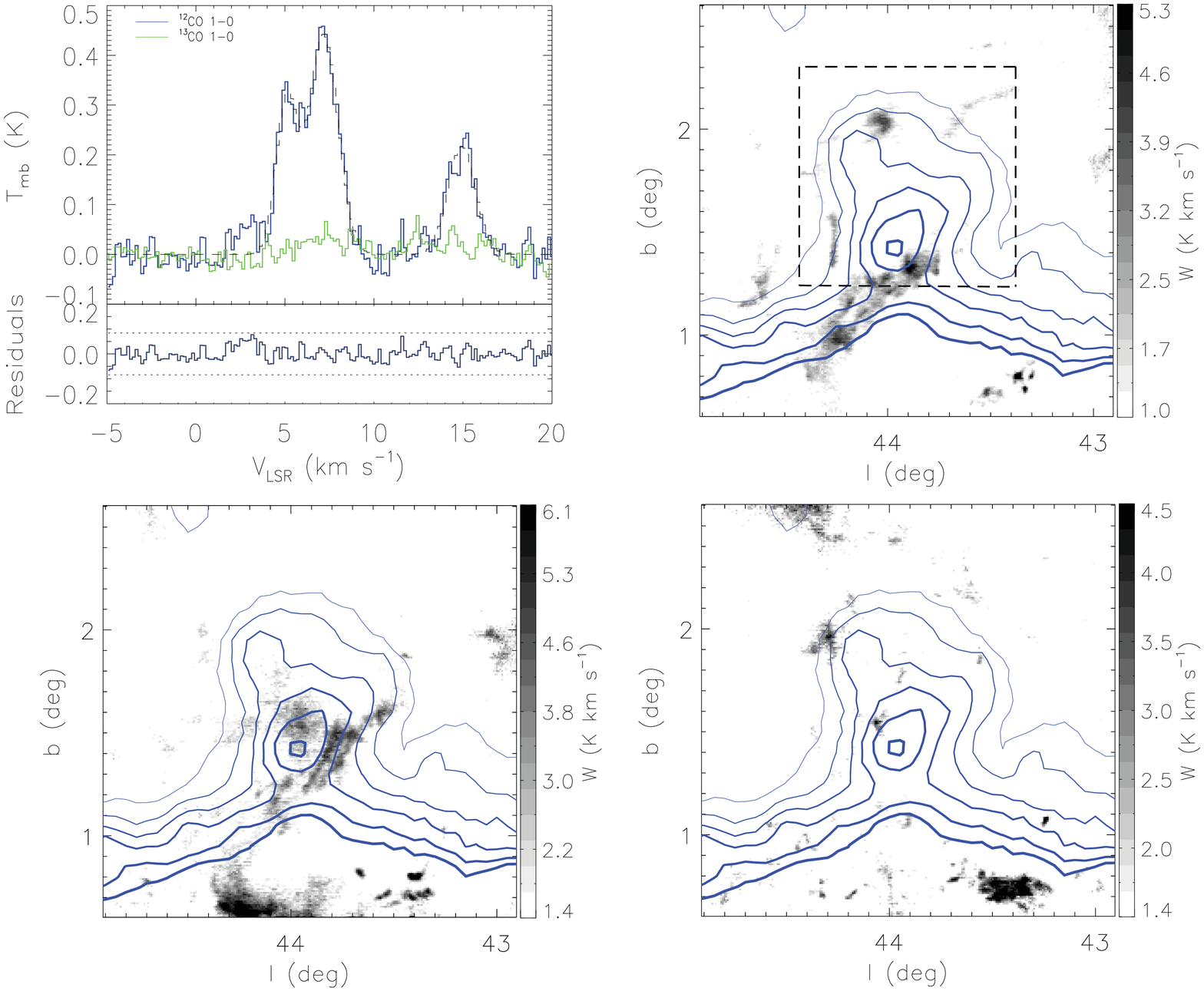,height=5.in,angle=0, clip=}
\hfil\hfil}}
\caption{{\sl upper-left:} Average \twCO~(J=1--0) (blue) and \thCO~(J=1--0) (green) spectra extracted from the SNR region defined in {\sl upper-right} pannel, with the best-fit Gaussian model and residuals of \twCO~(J=1--0) spectrum (see the fitted parameters in Table~\ref{tab:v5}).
3$\sigma$ level of residuals is shown by black dashed lines.
{\sl other three panels:} \twCO~(J=1--0) intensity maps integrated over the velocity ranges of 4.5--6~\km\ps\ ({\sl upper-right}), 6--9~\km\ps\ ({\sl lower-left}), and 13.5--16.5~\km\ps\ ({\sl lower-right}), overlaid with the same 6~cm radio continuum contours as in Figure~\ref{f:overall}.
The minimum values of \twCO\ emissions as shown in the colorbars are at $3\sigma$ values. 
}
\label{f:v5map}
\end{figure*}

\begin{table*}
\small
\begin{center}
\caption{Fitted parameters for the MCs in the velocity range of 0--20~\km\ps\ in the SNR region.\label{tab:v5}}
\begin{tabular}{cccccc}
\tableline\tableline
 Component & Line & Peak \tmb \tablenotemark{[1]} & Center \vlsr \tablenotemark{[2]} & FWHM \\
 & & (K)& (\km\ps) & (km~s$^{-1}$) \\
 \tableline
v5&\twCO\ (J=1-0) &$0.31\pm0.02$ &$5.10\pm0.05$ &$1.4\pm0.1$\\
&\thCO\ (J=1-0) &$\le0.02\tablenotemark{[3]}$ & -& -\\
\tableline
v7&\twCO\ (J=1-0) &$0.44\pm0.02$ &$7.18\pm0.04$ &$1.9\pm0.1$\\
&\thCO\ (J=1-0) & $\le0.02\tablenotemark{[3]}$ & -& -\\
\tableline
v15&\twCO\ (J=1-0) &$0.22\pm0.02$ &$14.93\pm0.05$ &$1.8\pm0.2$\\
&\thCO\ (J=1-0) & $\le0.02\tablenotemark{[3]}$ & -& -\\
 \tableline\tableline
\end{tabular}
\tablenotetext{[1]}{\tmb\ is the brightness temperature, and is corrected for beam efficiency using \tmb=T$_{\rm A}^{*}/\eta_{\rm mb}$.}
\tablenotetext{[2]}{\vlsr\ is the velocity with respect to the local standard of rest.}
\tablenotetext{[3]}{No \thCO~(J=1--0) emission visible, where we use the value of RMS as an upper limit.}
\end{center}
\end{table*}

As shown in Figure~\ref{f:v5map}, \twCO~(J=1--0) emission in the SNR region has three velocity components, i.e.\ 5, 7, and 15~\km\ps, in the velocity range of 0--20~\km\ps. 
These velocity components are probably from local MCs, i.e.\ parts of Serpens/Aquila molecular complex.
Nevertheless, the velocity range and the spatial distribution of the 5 and 7 \km\ps\ velocity components overlap each other, therefore, they are probably different parts of one MC.
The molecular gases of all the three velocity components distributes both inside and outside of the remnant.
Each emission line of the three velocity components in the SNR region can be fitted by one Gaussian component (see Figure~\ref{f:v5map}). The fitting parameters are listed in Table~\ref{tab:v5}.
We have not found any dynamical evidence of disturbance for these three velocity components.
There are also some \thCO~(J=1--0) emissions from the center of some clumps. No significant \CeiO~(J=1--0) emission was detected in this velocity range.   

The spatial distribution of the 5~\km\ps\ velocity component shows some correlations with the SNR~\snr, which distributes around the eastern and southern edge of the SNR.
Moreover, the 7~\km\ps\ velocity component distributes around the southwestern border of the SNR, with also a clump located near the radio peak of the remnant.
No spatial correlation is found between the 15~\km\ps\ velocity component and the remnant.
Note that the spatial correlations can be just chance correlations.

\subsection{The $\sim$50~\km\ps\ Velocity Component}\label{subsec:50vc}
\begin{figure*}[ptbh!]
\centerline{{\hfil\hfil
\psfig{figure=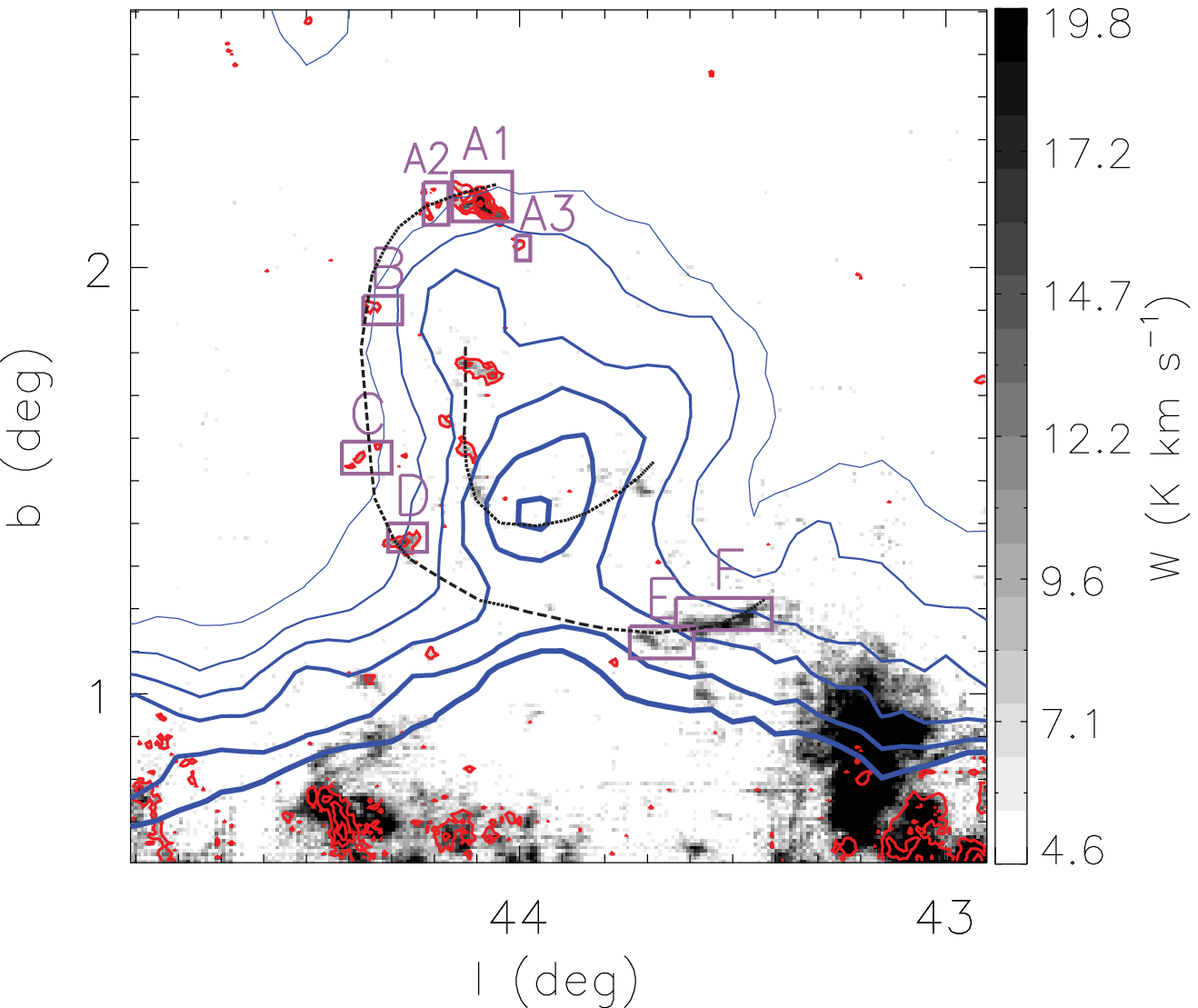,height=5.5in,angle=0, clip=}
\hfil\hfil}}
\caption{
Intensity map of \twCO~(J=1--0) emission in the velocity range of 40--70~\km\ps, overlaid with the same 6~cm radio continuum contours as in Figure~\ref{f:overall}.
\twCO~(J=1--0) emission in the velocity range of 40--55~\km\ps\ are also shown by red contours.
The contour levels of the \twCO\ emission begin at $5\sigma$ and also increase by $5\sigma$, where the $\sigma$ value is 1.1~K~\km\ps.
Two groups of molecular clumps are sketched by two black dashed and dotted lines, which are distributed around the remnant's border and radio shell peak.
The purple regions are selected for spectral analysis.
}
\label{f:v50map}
\end{figure*}

The spatial distribution of the $\sim$50~\km\ps\ velocity component shows correlation with the SNR~\snr, which contains a series of small molecular clumps distributing along the radio shell of the remnant. 
As sketched by black dashed and dotted lines in Figure~\ref{f:v50map}, the clumps can be divided into two groups, one distributes around the border of the remnant, enclosing the remnant's eastern half, while, the other group distributes around the remnant's radio shell peak. 
Note that the southern border of the remnant is not very clear, and regions E and F may be not associated with regions A, B, C, and D.

\begin{figure*}[ptbh!]
\centerline{{\hfil\hfil
\psfig{figure=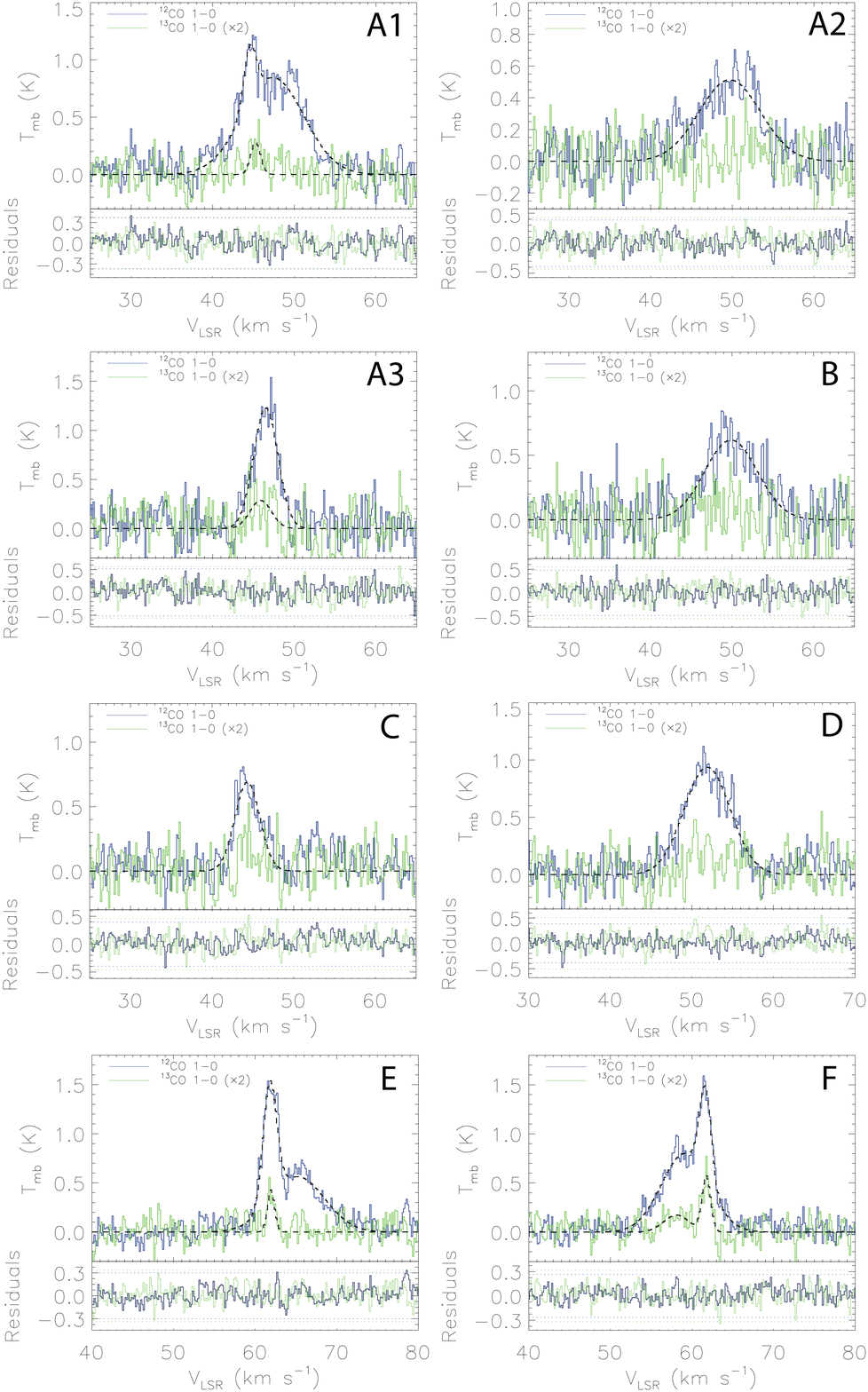,height=6.5in,angle=0, clip=}
\hfil\hfil}}
\caption{CO spectra extracted from the selected regions shown in Figure~\ref{f:v50map}, together with their best-fit Gaussian model and residuals. Masks of integrated intensities larger than $3\sigma$ are applied in extracting spectra, over the velocity ranges of 40--55\km\ps\ for the regions of A1--D and 55--70\km\ps\ for the regions of E and F.
The fitted parameters are listed in Table~\ref{tab:v50}.
3$\sigma$ levels of the residuals are also shown by blue and green dotted lines for \twCO\ and \thCO, respectively.
The \thCO~(J=1--0) spectra as well as their fitting models and residuals are multiplied by a factor of 2 for better visibility.
}
\label{f:spec}
\end{figure*}

\begin{table*}
\begin{center}
\caption{The same as Table~\ref{tab:v5}, but for other velocity components.\label{tab:v50}}
\begin{tabular}{llcccccc}
\tableline\tableline
region&component&Line&Peak \tmb&Center \vlsr&FWHM\\
&&&(K)&(\km\ps)&(\km\ps)\\
\tableline
A1&narrow&\twCO~(J=1--0)&0.49$\pm{0.06}$&44.55$\pm{0.10}$&1.8$\pm{0.3}$\\
&&\thCO~(J=1--0)&0.14$\pm{0.03}$&45.2$\pm{0.3}$&1.4$\pm{0.9}$\\
&broad&\twCO~(J=1--0)&0.84$\pm{0.03}$&47.4$\pm{0.2}$&8.9$\pm{0.3}$\\
&&\thCO~(J=1--0)&$\leq{0.07}$&-&-\\
\tableline
A2&broad&\twCO~(J=1--0)&0.51$\pm{0.03}$&49.7$\pm{0.3}$&8.9$\pm{0.5}$\\
&&\thCO~(J=1--0)&$\leq{0.07}$&-&-\\
\tableline
A3&broad&\twCO~(J=1--0)&1.23$\pm{0.06}$&46.49$\pm{0.08}$&3.5$\pm{0.2}$\\
&&\thCO~(J=1--0)&0.14$\pm{0.04}$&45.9$\pm{0.4}$&3.1$\pm{0.9}$\\
\tableline
B&broad&\twCO~(J=1--0)&0.62$\pm{0.04}$&49.9$\pm{0.2}$&7.6$\pm{0.5}$\\
&&\thCO~(J=1--0)&$\leq{0.08}$&-&-\\
\tableline
C&broad&\twCO~(J=1--0)&0.69$\pm{0.04}$&44.31$\pm{0.10}$&3.6$\pm{0.3}$\\
&&\thCO~(J=1--0)&$\leq{0.07}$&-&-\\
\tableline
D&broad&\twCO~(J=1--0)&0.94$\pm{0.03}$&51.9$\pm{0.1}$&6.7$\pm{0.3}$\\
&&\thCO~(J=1--0)&$\leq{0.07}$&-&-\\
\tableline
E&narrow&\twCO~(J=1--0)&1.20$\pm{0.05}$&61.81$\pm{0.04}$&2.0$\pm{0.1}$\\
&&\thCO~(J=1--0)&0.21$\pm{0.04}$&61.94$\pm{0.08}$&1.0$\pm{0.2}$\\
&broad&\twCO~(J=1--0)&0.56$\pm{0.03}$&65.2$\pm{0.3}$&7.8$\pm{0.4}$\\
&&\thCO~(J=1--0)&$\leq{0.06}$&-&-\\
\tableline
F&narrow&\twCO~(J=1--0)&0.94$\pm{0.05}$&61.65$\pm{0.04}$&1.7$\pm{0.1}$\\
&&\thCO~(J=1--0)&0.28$\pm{0.03}$&61.75$\pm{0.07}$&1.2$\pm{0.2}$\\
&broad&\twCO~(J=1--0)&0.80$\pm{0.03}$&59.1$\pm{0.2}$&6.8$\pm{0.3}$\\
&&\thCO~(J=1--0)&0.09$\pm{0.02}$&58.1$\pm{0.4}$&3.7$\pm{1.0}$\\
\tableline\tableline
\end{tabular}
\end{center}
\end{table*}

The CO lines of $\sim$50~\km\ps\ velocity component extracted from the molecular clumps are much broader than the lines in the velocity range of 0--20~\km\ps\ (see Figure~\ref{f:spec}).
These broad lines are not combinations of multiple narrow lines, which indicate the existence of disturbances in the corresponding MCs. 
We have fitted the CO lines with Gaussian functions. The fitting parameters are listed in Table~\ref{tab:v50}. Most of the lines in the velocity range of 40--70~\km\ps\ can be fitted by one broad Gaussian component, except the lines from region A1, E, and F, which contain a narrow Gaussian component associated with a broad component. 
The molecular clump in region A1 is one of the largest clumps, probably associated with the molecular gases in region A2 and A3. There is \thCO\ (J=1--0) emission corresponding to the narrow component of \twCO\ (J=1--0) emission in region A1.
MCs in regions E and F are at the velocity of $\sim$60~\km\ps\ other than $\sim$50~\km\ps\ for the other MCs, and are located in the south-west. 
The shapes of these two MCs are arc-like, which seem to be lying around the edge of the remnant.

\begin{figure*}[ptbh!]
\centerline{{\hfil\hfil
\psfig{figure=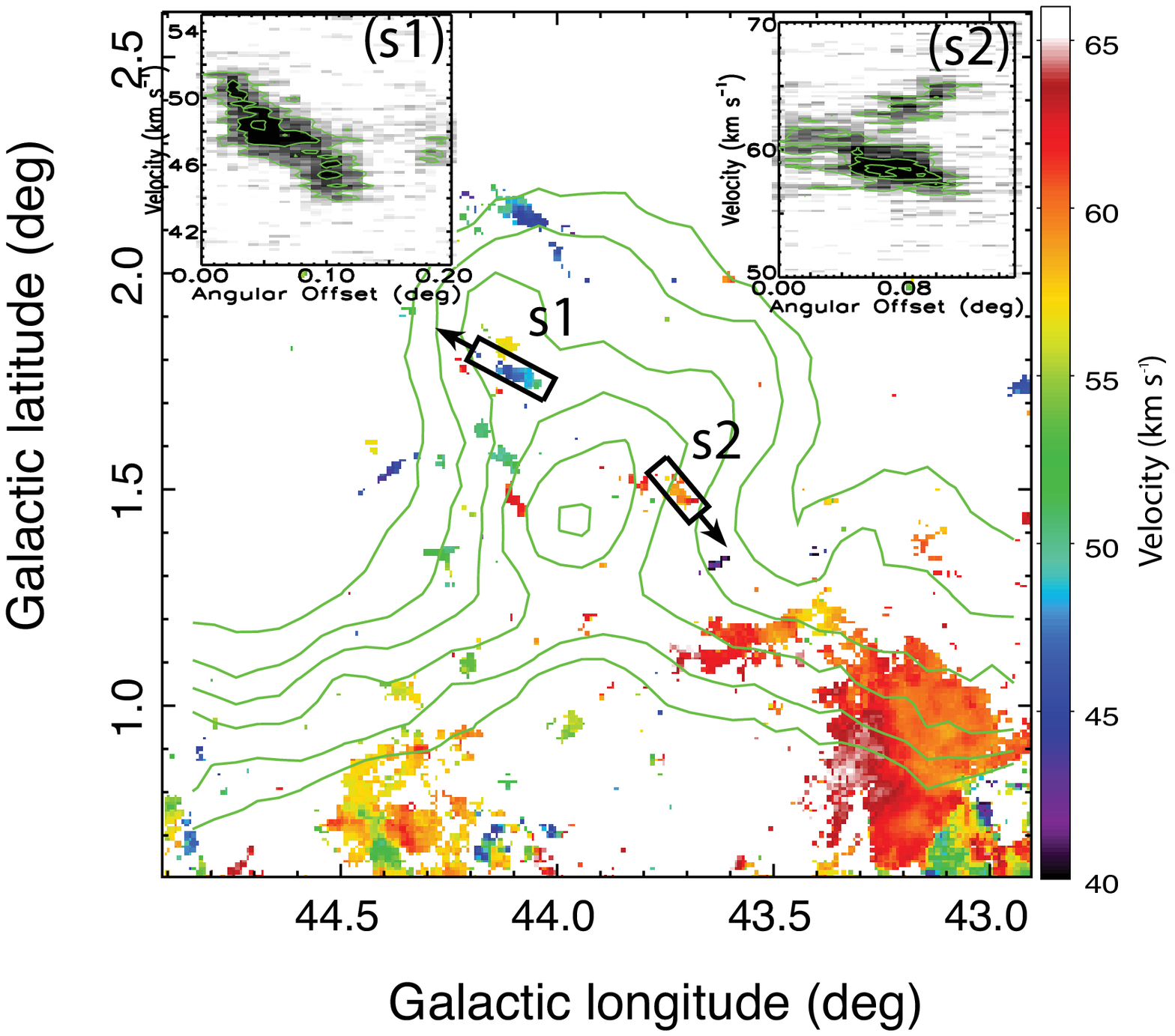,height=5.5in,angle=0, clip=}
\hfil\hfil}}
\caption{
Intensity weighted mean velocity (1st moment) map of \twCO\ emission in the velocity range of 40--55~\km\ps, overlaid with the same 6~cm radio continuum contours as in Figure~\ref{f:v5map}. Position-velocity maps of \twCO\ along the strips indicated by black rectangular regions are also presented, where the contour levels are from $3\sigma$ and in a step of $3\sigma$, where the $\sigma$ value is 0.3~K. 
}
\label{f:pv50}
\end{figure*}

There are two molecular strips nearby the remnant's radio shell peak, which show velocity gradients in position-velocity maps (see Figure~\ref{f:pv50}).
The velocity extents of the strips are similar to the width of the broad emission lines in the other regions, which indicates the same origin of disturbance, probably being the remnant's shock. 
Using the velocities of the nearby quiet narrow components as reference velocities, i.e.\ $\sim$45~\km\ps\ in region A1 for strip s1 and $\sim$62~\km\ps\ in region E and F for strip s2, the molecular emissions in both regions are red-shifted. Therefore, the two molecular strips are probably located at the far side of the remnant.

\section{Discussion}\label{sec:discuss}
\subsection{The Local Molecular Component}\label{subsec:v5mc}
We estimate the near and far kinematic distances of the $\sim$5~\km\ps\ MC as 0.31$\pm{0.06}$~kpc and 11.6$\pm{0.6}$~kpc, respectively, based on a full distance probability density function\footnote{http://bessel.vlbi-astrometry.org/node/378} \citep{Reid+2016,Reid+2019}. 
The systemic velocity of the MC is estimated as the mean velocity of the v5 and v7 components (see Table~\ref{tab:v5}), and its uncertainty as the difference between the velocities of the two components.
The physical parameters of these local velocity components are estimated using the method described in detail in \cite{Zhou+2016} (see Table~\ref{tab:phy}).
The area beam-filling factors of both \twCO~(J=1--0) and \thCO~(J=1--0) emissions are assumed to be the same, which are calculated by the ratio of the number of points with a detected \twCO~(J=1--0) emission to the total number of points in the SNR region.
The column densities of the $\sim$5~\km\ps\ MCs are small, with \twCO~(J=1--0) emission be not optically thick, hence, the excitation temperatures can not be well confined. 
If the $\sim5$~\km\ps\ MC is at the near distance, the masses of molecular gases in the SNR region are less than $\sim10^{2}\dua^{2}~\Msun$ (see Table~\ref{tab:phy}), which are over one order of magnitude smaller than their virial masses. It indicates that the MCs are not stable. However, the filamentary shape cloud may be not confined by gravity but by magnetic field instead \citep{Contreras+2013}. If the MC is at the far distance, the virial mass over mass ratios will be $d/0.31~pc\simeq37$ times smaller, i.e.\ about 1, then the cloud would be in virial equilibrium.
Su (2019, private communication) stated that the $\sim$5~\km\ps\ MC is located at a near distance based on the $Gaia$ data, i.e.\ about 240~pc.

If the SNR is associated with the $\sim$5~\km\ps\ MC and locates at the near distance of $0.31\pm{0.06}$~kpc, its radius would be $\sim$2.7$\dua$~pc, where $\dua$ is the ratio between the distance of the remnant and the distance of 0.31~kpc.
Assuming the lengths in the line of sight (LOS, represented by $l$) are comparable to the widths of MCs, particularly for MCs outside of the SNR, we could derive the number density as $n_{\rm cloud}={\rm N}({\rm H}_{2})/l\sim10^{20}~{\rm cm}^{-2}/0.5$~pc$\sim60\dua^{-1}$~cm$^{3}$.
The possible molecular shock velocity could be estimated as $v_{\rm cloud}=4v_{\rm shift}/3\sim2.7$~km~s$^{-1}$, where $v_{\rm shift}$ is the difference of center \vlsr\ between 5~\km\ps\ and 7~\km\ps\ velocity components. 
Using the density of the ambient interstellar medium as the Galactic fiducial density $n_0=1~\cm^{-3}$, we get the density contrast as $\chi=n_{\rm cloud}/n_0\sim\beta v_{\rm s}^2/v_{\rm cloud}^2\sim$60$\dua^{-1}$, where $v_{\rm s}$ is the velocity of the remnant's forward shock, and the constant $\beta$ is adopted as unity \citep{McKeeCowie1975,Orlando+2005}.
Accordingly, we get the forward shock velocity as $v_{\rm s}$=21$\dua^{-0.5}$~\km\ps\ and the post-shock temperature as 5.9$\E{3}\dua^{-1}$~K. Such low forward shock velocity and post-shock temperature indicate that the remnant would be in the late radiative phase. 
Consequently, the age of the remnant could be estimated as $t=2r_{\rm s}/(7v_{\rm s})\sim$9.9$\E{4}\dua^{1.5}~\yr$ \citep{McKeeOstriker1977}, and the explosion energy \\
$E=6.8\E{43}n_{0}^{1.16}\frac{v_{\rm s}}{1~{\rm km~s}^{-1}}^{1.35}\frac{r_{\rm s}}{1~{\rm pc}}^{3.16}\zeta_{\rm m}^{0.161}~{\rm erg} \\
\sim4.9\E{47}\dua^{2.485}~{\rm erg}$, \\
where $\zeta_{\rm m}={\rm Z/Z}_\odot$ is the metallicity parameter, which is set to 1 \citep{Cioffi+1988}.
The explosion energy would be four orders of magnitude lower than the canonical value ($10^{51}$~erg), which is significantly low even for a sub-energetic core-collapse supernova explosion, where $E_{\rm SN}\sim10^{49.5}$~erg \citep{Pastorello+2004, Chevalier2005, Zhou+2014}.
It is also possible that the SNR is evolving in a low-density environment, e.g.\ inside a wind-blown bubble. 
However, we have the explosion energy $E\propto n_{0}^{1.835}$, and we would get a lower explosion energy for a lower ambient density, e.g.\ $E\sim2.1\E{46}\dua^{2.485}$~erg for the ambient density of $n_0=0.01~\cm^{-3}$ \citep{Dwarkadas2007, ChoKang2008, Renaud+2010}. 
If the SNR is at the far distance, i.e.\ 11.6~kpc, the derived density of the molecular gas is $\sim1.6~\cm^{-3}$ that is too low for an MC (typical mean density $\sim100~\cm^{-3}$).
Besides, the shock speed and the remnant age would be 3.4~\km\ps\ and 8.4$\E{6}~\yr$, respectively. With the speed of shock less than that of interstellar turbulence \citep[about 10~\km\ps;][]{Spitzer1978} and the large age, the remnant would be essentially invisible in the radio continuum \citep[e.g.][]{StilIrwin2001}.
Indeed, the derived age is larger than that of any SNR known \citep[e.g.][]{StilIrwin2001, Koo+2006, XiaoZhu2014}.
The absence of dynamical evidence as well as the unreasonable derived parameters suggest that the SNR is not associated with the $\sim$5~\km\ps\ MC. Therefore, the spatial correlation between them is just a chance correlation.

\begin{table*}
\begin{center}
\caption{The Derived Physical Parameters for different MCs.\label{tab:phy}}
\begin{tabular}{llcccccc}
\tableline\tableline
region&component&\tex \tablenotemark{[1]}&$\tau$(\thCO) \tablenotemark{[1]}&\ncol \tablenotemark{[2]}&$M$ \tablenotemark{[2]}&$M_{\rm vir}$ \tablenotemark{[3]}\\
&&(K)&&(10$^{20}$ cm$^{-2}$)&($10^{2}~\Msun$)&($10^{2}~\Msun$)\\
\tableline
\multicolumn{6}{c}{\sl The Local Molecular Component}\\
\tableline 
SNR & v5 & $>3.6$ & $<0.07$ & $<1.0$ (2.0) & $<0.3\dua^2$ ($0.4\dua^2$)&18$\dua$\\
SNR & v7 & $>3.6$ & $<0.05$ & $<0.9$ (2.6) & $<0.3\dua^2$ ($0.8\dua^2$)&34$\dua$\\
SNR & v15 & $>3.4$ & $<0.1$ & $<1.8$ (2.0) & $<0.4\dua^2$ ($0.4\dua^2$)&30$\dua$\\
\tableline\tableline
\multicolumn{6}{c}{\sl The $\sim$50~\km\ps\ Molecular Component}\\
\tableline
A1&narrow&3.4&0.3&4.5 (1.7)&$2.3 {\du}^2$ ($0.87 \du^2$)&46$\du$ \\
&broad&$>$3.8&$<0.09$&$<8.5$ (14)&$<4.4\du^2$ (7.4$\du^2$)&1100$\du$ \\
\tableline
A2&broad&$>$3.3&$<$0.2&$<$13 (8.7)&$<2.9{\du}^2$ (1.9$\du^2$) &740$\du$ \\
\tableline
A3&broad&4.3&0.1&4.8 (8.2)&0.34${\du}^2$ (0.59$\du^2$) &77$\du$ \\
\tableline
B&broad&$>$3.5&$<0.2$&$<11 $(9.0)&$<0.42{\du}^2$ (0.35$\du^2$) &540$\du$ \\
\tableline
C&broad&$>$3.6&$<0.2$&$<4.0$ (4.8)&$<0.88{\du}^2$ (1.1$\du^2$) &160$\du$  \\
\tableline
D&broad&$>$3.9&$<0.08$&$<5.9$ (12)&$<1.4{\du}^2$ ($2.7\du^2$) &420$\du$  \\
\tableline
E&narrow&4.2&$0.2$&$2.4$ (4.6)&$1.3{\du}^2$ ($2.5\du^2$) &53$\du$  \\
&broad&$>$3.4&$<0.2$&$<8.7$ (8.4)&$<4.7{\du}^2$ ($4.5\du^2$) &810$\du$  \\
\tableline
F&narrow&3.9&0.3&4.8 (3.1)&$4.3{\du}^2$ ($2.7\du^2$) &55$\du$  \\
&broad&3.7&$0.1$&$4.8$ (10)&$4.2{\du}^2$ ($9.2\du^2$) &890$\du$  \\
\tableline\tableline
\end{tabular}
\tablenotetext{[1]}{Using the assumption of local thermal equilibrium (LTE). 
See the details of calculation method in \cite{Zhou+2016}.}
\tablenotetext{[2]}{Derived from \thCO\ column density by assuming the \thCO\ abundance of 1.4$\E{-6}$ \citep{Ripple+2013}. For comparison, we also show the values in the brackets, which are estimated by using the conversion factor $N$(H$_2)/{W}(^{12}{\rm CO)}\simeq1.8\times10^{20}~{\rm cm}^{-2}~{\rm K}^{-1}~{\rm km}^{-1}$~s \citep{Dame+2001}. $\dua$ and $\du$ stands for $d/(0.31~{\rm kpc})$ and $d/(3.1~{\rm kpc})$, respectively, where d is the distance to the SNR~\snr\ in unit of kpc.}
\tablenotetext{[3]}{Calculated by ${k}\times{L}\times\Delta v^{2}$, where $L$ is the size of the region, and $\Delta v$ is the velocity width (FWHM) of \twCO~(J=1--0), considering the elongated shape of the MCs and assuming a constant density distribution, $k$ is set to 173 \citep{MacLaren+1988, Ren+2014}.}
\end{center}
\end{table*}

\subsection{The $\sim$50~\km\ps\ Molecular Component}\label{subsec:v50mc}
Applying the full distance probability density function as the above, we get the near and far distances of the $\sim$50~\km\ps\ velocity component to be 3.1$\pm{1.2}$~kpc and 8.5$\pm{1.3}$~kpc, respectively. The systemic velocity of the MC is estimated as the mean velocity of the components from all the A1 to F regions, and its uncertainty as the largest velocity difference between these components.
In regions A1, E and F, we use the velocities of the narrow components. 
If we exclude regions E and F, we get the near and far distances to be 2.8$\pm{0.7}$~kpc and 8.9$\pm{0.8}$~kpc, respectively. The two pairs of distances are consistent in their error ranges.
Using the same calculation method, we also estimate the physical parameters of the $\sim$50~\km\ps\ velocity component, which are listed in Table~\ref{tab:phy}.
The $\sim$50~\km\ps\ velocity component comprises a series of small molecular clumps, which are distributed along the radio border and the radio shell peak of the remnant.
Broad CO lines are detected in all the selected regions, however, those with associated narrow CO lines are only detected in the regions that contain large molecular clumps, i.e.\ A1, E, and F (see Figures~\ref{f:v50map} and \ref{f:spec}).
Due to the small optical depths of CO emissions, the excitation temperatures of these molecular clumps can not be well confined. 
The column densities as well as the masses of the disturbed molecular gases can not be well estimated by using the conversion factor \ncol/$W$(\twCO). 
The distance factor $\du$ would be 1 or 2.7 for the $\sim50$~\km\ps\ MC being at the near or far distance, respectively. It will not change the ratios between virial mass and mass in magnitude.
All the masses of the broad components are about two orders of magnitude smaller than their virial masses, which confirms the existence of strong disturbances in the clumps. It indicates that the broad components may be from molecular gases shocked by the remnant's blast wave.
The masses of the narrow components are also about one order of magnitude smaller than their virial masses, which indicates that the quiet molecular gases in these regions can not be confined by gravity too.
A large percent of molecular gases in the large molecular clumps, in regions A1, E and F, are probably shocked, with the masses of the disturbed molecular gases comparable to that of the quiet molecular gases. 
Moreover, in the small clumps in regions A2--D, there may be no quiet molecular gas left, with all molecular gases disturbed.

Two molecular strips with velocity gradients are also detected around the remnant's radio shell peak (see Figure~\ref{f:pv50}).
Their velocity extents are at the same level as the width of the broad emission lines in the other regions, which probably originate from the remnant's shock too.
The velocity gradients of the strips are both $\sim$1.7$\du^{-1}$~\km\ps pc$^{-1}$.
Strip s1 is more straight than strip s2, which is likely to have been shocked from one side to another side. The case of strip s2 is not that simple, and it has a complicated pattern in position-velocity diagram. Strip s2 is curved, and may be shocked on one side first then on the other side before engulfed in the remnant.
Considering that the molecular shocks has not passed through the strips,
we estimate the dynamical timescales of the strips as $\tau_{\rm strip}\lesssim l_{\rm strip}/\Delta v_{\rm strip}\sim$6$\E{5}\du$~yr, where $l_{\rm strip}$ and $\Delta v_{\rm strip}$ are the extents and the velocity spans of the strips, respectively.

If the remnant is at the near distance of the $\sim$50~\km\ps\ velocity component, i.e.\ $\sim$3.1~kpc, its radius is $\sim$27~pc. 
We consider region A1 as a representative region, and its length in the LOS is assumed to be comparable to its size $l\sim0\degree.05$. Accordingly, we have the number density of the clump as $n_{cloud}\sim$54$\du^{-1}~\cm^{-3}$.
Applying the calculation method used above, we estimate the velocity of the remnant's forward shock as $v_s\sim$29$\du^{-0.5}$~\km\ps\ and the post-shock temperature as 1.2$\E{4}\du^{-1}$~K, which indicates that the remnant is in the radiative phase.
Therefore, we get the age of the remnant as $t\sim$2.6$\E{5}\du^{1.5}$~yr and the explosion energy as $E\sim$2.2$\E{50}\du^{2.485}$~erg.
The age estimated for the remnant is consistent with the dynamical timescales of the strips s1 and s2. 
It is also possible that the remnant locates at the far distance as $\sim$8.5~kpc.
In this case, we can get the age of the remnant as $\sim1.2\E{6}$~yr by applying the distance factor $\du$ as 2.7. 
The estimated age is comparable to that of the oldest radio detected SNR, i.e.\ G55.0+0.3 \citep[$\sim1.9\E{6}$~yr;][]{Matthews+1998}. 
However, the radio continuum emission of \snr\ is much brighter, with the 1~GHz surface brightness of $\Sigma_{1~GHz}\sim$4$\E{-22}$~\brightnessunit \citep{Reich+1988} more than five times higher than that of G55.0+0.3 ($\Sigma_{1~GHz}\sim$7$\E{-23}$~\brightnessunit). 
The slow shock of such old SNR is considered to be inefficient to accelerate particles to relativistic energies responsible for radio synchrotron emission \citep{DraineMcKee1993}.
Actually, SNRs at the age around one million years are more likely detected by \HI\ observations, e.g.\ GSH~138-01-94 \citep{StilIrwin2001} and GSH~90-28-17\citep{XiaoZhu2014}, and these SNRs are indeed very weak in radio continuum emission \citep[e.g.\ $\Sigma_{1~GHz}\lesssim7\E{-23}$~\brightnessunit;][]{Koo+2006,XiaoZhu2014}.

Therefore, by both dynamical evidence and spatial correlation, we confirm that the $\sim$50~\km\ps\ MC is associated with the SNR~\snr.
As implied by 
the bright radio continuum emission, the SNR is probably at the near distance as $\sim3.1$~kpc.
The total mass of shocked molecular gases is $\sim400~\Msun$ in regions A3 and F, and is less than $1700~\Msun$ in all the selected regions. Accordingly, the total kinematic energy of shocked molecular gases in the selected regions is in the range of $\sim3.6\E{46}$~erg to $\sim1.5\E{47}$~erg.
There are also some molecular gases left in the remnant but outside the selected regions, however, they would be no more than the molecular gases in the selected regions.

\section{Summary}\label{sec:summary}
We have studied the SNR~\snr\ to investigate its molecular environment. Correlations between the SNR and MCs at different velocities are examined, based on both spatial distribution and dynamical evidence of CO line emissions.
We found that the spatial distributions of both the $\sim$5~\km\ps\ and $\sim$50~\km\ps\ velocity components show some correlations with the remnant.
However, no dynamical evidence of disturbance was found for the $\sim$5~\km\ps\ velocity component.
At the distance of the $\sim$5~\km\ps\ velocity component, either near or far distance, the derived physical parameters are unreasonable too.
We conclude that the SNR is not associated with the $\sim$5~\km\ps\ velocity component, and their spatial correlation is just a chance correlation.

For the $\sim$50~\km\ps\ velocity component, dynamical evidence of disturbances, as well as the spatial correlation, indicate that the MC is associated with the SNR.
The remnant is probably at the near kinematic distance of the $\sim$50~\km\ps\ velocity component, i.e.\ $\sim3.1$~kpc, as implied by its bright radio continuum emission.
Accordingly, we obtained the age of the remnant as about $2.6\E{5}$~yr and the explosion energy as about $2.2\E{50}$~erg.
All of the CO spectra extracted from the molecular clumps distributed along the border of the remnant are with broad components presented, which can be fitted by Gaussian functions. 
By further spectral analysis, we get the total mass of the shocked molecular gases in these molecular clumps to be in the range of $\sim$400--$1700~\Msun$, and the total kinetic energy in the range of $\sim3.6\E{46}$--$1.5\E{47}$~erg.
Velocity gradients were also detected along two molecular strips around the remnant's radio shell peak, which probably locate at the far side of the remnant.

\acknowledgments
We are grateful to all the members in the Milky Way Scroll Painting-CO line survey group, especially the staff of Qinghai Radio Observing Station at Delingha for the support during the observation.
We thank the anonymous referee for providing very helpful comments that improved the paper and its conclusions.
MWISP project is supported by the National Key R\&D Program of China grant no.\ 2017YFA0402700 and the Key Research Program of Frontier Sciences, CAS, grant no.\ QYZDJ-SSW-SLH047.
Y.C. acknowledges support from the Key R\&D Program grants 2017YFA0402600 and 2015CB857100.
This work is supported by the The National R\&D Infrastructure and Facility Development Program of China, "Fundamental Science Data Sharing Platform" (DKA2017-12-02-08), and the NSFC grants 11403104, 11773014, 11633007, and 11851305.

\bibliographystyle{apj}
\bibliography{ms.bbl}

\begin{thebibliography}{61}
\expandafter\ifx\csname natexlab\endcsname\relax\def\natexlab#1{#1}\fi

\bibitem[{{Aharonian} \& {Atoyan}(1996)}]{AharonianAtoyan1996}
{Aharonian}, F.~A., \& {Atoyan}, A.~M. 1996, \aap, 309, 917

\bibitem[{{Chen} {et~al.}(2017){Chen}, {Xiong}, \& {Yang}}]{Chen+2017}
{Chen}, X., {Xiong}, F., \& {Yang}, J. 2017, \aap, 604, A13

\bibitem[{{Chen} {et~al.}(2014){Chen}, {Jiang}, {Zhou}, {Su}, {Zhou}, {Li}, \&
  {Zhang}}]{Chen+2014}
{Chen}, Y., {Jiang}, B., {Zhou}, P., {et~al.} 2014, in IAU Symposium, Vol. 296,
  Supernova Environmental Impacts, ed. A.~{Ray} \& R.~A. {McCray}, 170--177

\bibitem[{{Chevalier}(2005)}]{Chevalier2005}
{Chevalier}, R.~A. 2005, \apj, 619, 839

\bibitem[{{Cho} \& {Kang}(2008)}]{ChoKang2008}
{Cho}, H., \& {Kang}, H. 2008, NewA, 13, 163

\bibitem[{{Cioffi} {et~al.}(1988){Cioffi}, {McKee}, \&
  {Bertschinger}}]{Cioffi+1988}
{Cioffi}, D.~F., {McKee}, C.~F., \& {Bertschinger}, E. 1988, \apj, 334, 252

\bibitem[{{Contreras} {et~al.}(2013){Contreras}, {Rathborne}, \&
  {Garay}}]{Contreras+2013}
{Contreras}, Y., {Rathborne}, J., \& {Garay}, G. 2013, \mnras, 433, 251

\bibitem[{{Dame} {et~al.}(2001){Dame}, {Hartmann}, \& {Thaddeus}}]{Dame+2001}
{Dame}, T.~M., {Hartmann}, D., \& {Thaddeus}, P. 2001, \apj, 547, 792

\bibitem[{{de Wilt} {et~al.}(2017){de Wilt}, {Rowell}, {Walsh}, {Burton},
  {Rathborne}, {Fukui}, {Kawamura}, \& {Aharonian}}]{deWilt+2017}
{de Wilt}, P., {Rowell}, G., {Walsh}, A.~J., {et~al.} 2017, \mnras, 468, 2093

\bibitem[{{Dickman} {et~al.}(1992){Dickman}, {Snell}, {Ziurys}, \&
  {Huang}}]{Dickman+1992}
{Dickman}, R.~L., {Snell}, R.~L., {Ziurys}, L.~M., \& {Huang}, Y.-L. 1992,
  \apj, 400, 203

\bibitem[{{Draine} \& {McKee}(1993)}]{DraineMcKee1993}
{Draine}, B.~T., \& {McKee}, C.~F. 1993, \araa, 31, 373

\bibitem[{{Dwarkadas}(2007)}]{Dwarkadas2007}
{Dwarkadas}, V.~V. 2007, in Revista Mexicana de Astronomia y Astrofisica
  Conference Series, Vol.~30, Revista Mexicana de Astronomia y Astrofisica
  Conference Series, 49--56

\bibitem[{{Eger} {et~al.}(2011){Eger}, {Rowell}, {Kawamura}, {Fukui},
  {Rolland}, \& {Stegmann}}]{Eger+2011}
{Eger}, P., {Rowell}, G., {Kawamura}, A., {et~al.} 2011, \aap, 526, A82

\bibitem[{{Frail} {et~al.}(2013){Frail}, {Claussen}, \&
  {M{\'e}hault}}]{Frail+2013}
{Frail}, D.~A., {Claussen}, M.~J., \& {M{\'e}hault}, J. 2013, \apjl, 773, L19

\bibitem[{{Fukuda} {et~al.}(2014){Fukuda}, {Yoshiike}, {Sano}, {Torii},
  {Yamamoto}, {Acero}, \& {Fukui}}]{Fukuda+2014}
{Fukuda}, T., {Yoshiike}, S., {Sano}, H., {et~al.} 2014, \apj, 788, 94

\bibitem[{{Gaensler} {et~al.}(2008){Gaensler}, {Tanna}, {Slane}, {Brogan},
  {Gelfand}, {McClure-Griffiths}, {Camilo}, {Ng}, \& {Miller}}]{Gaensler+2008}
{Gaensler}, B.~M., {Tanna}, A., {Slane}, P.~O., {et~al.} 2008, \apjl, 680, L37

\bibitem[{{Green}(2019)}]{Green2019}
{Green}, D.~A. 2019, arXiv e-prints, arXiv:1907.02638

\bibitem[{{Hewitt} {et~al.}(2009){Hewitt}, {Yusef-Zadeh}, \&
  {Wardle}}]{Hewitt+2009}
{Hewitt}, J.~W., {Yusef-Zadeh}, F., \& {Wardle}, M. 2009, \apjl, 706, L270

\bibitem[{{Jeong} {et~al.}(2012){Jeong}, {Byun}, {Koo}, {Lee}, {Lee}, \&
  {Kang}}]{Jeong+2012}
{Jeong}, I.-G., {Byun}, D.-Y., {Koo}, B.-C., {et~al.} 2012, \apss, 342, 389

\bibitem[{Jiang {et~al.}(2010)Jiang, Chen, Wang, Su, Zhou, Safi-Harb, \&
  DeLaney}]{Jiang+2010}
Jiang, B., Chen, Y., Wang, J., {et~al.} 2010, ApJ, 712, 1147

\bibitem[{{Kilpatrick} {et~al.}(2016){Kilpatrick}, {Bieging}, \&
  {Rieke}}]{Kilpatrick+2016}
{Kilpatrick}, C.~D., {Bieging}, J.~H., \& {Rieke}, G.~H. 2016, \apj, 816, 1

\bibitem[{{Koo} {et~al.}(2006){Koo}, {Kang}, \& {Salter}}]{Koo+2006}
{Koo}, B.-C., {Kang}, J.-h., \& {Salter}, C.~J. 2006, \apjl, 643, L49

\bibitem[{{Lau} {et~al.}(2017){Lau}, {Rowell}, {Burton}, {Fukui}, {Aharonian},
  {Oya}, {Vink}, {Ohm}, \& {Casanova}}]{Lau+2017}
{Lau}, J.~C., {Rowell}, G., {Burton}, M.~G., {et~al.} 2017, \mnras, 464, 3757

\bibitem[{{Li} \& {Chen}(2012)}]{LiChen2012}
{Li}, H., \& {Chen}, Y. 2012, \mnras, 421, 935

\bibitem[{{Liu} {et~al.}(2017){Liu}, {Chen}, {Zhang}, {Liu}, {He}, {Zhou},
  {Zhou}, \& {Su}}]{Liu+2017}
{Liu}, B., {Chen}, Y., {Zhang}, X., {et~al.} 2017, \apj, 851, 37

\bibitem[{{Liu} {et~al.}(2018){Liu}, {Chen}, {Chen}, {Zhou}, {Wang}, \&
  {Su}}]{Liu+2018}
{Liu}, Q.-C., {Chen}, Y., {Chen}, B.-Q., {et~al.} 2018, \apj, 859, 173

\bibitem[{{Ma} {et~al.}(2019){Ma}, {Wang}, {Zhang}, {Li}, \& {Yang}}]{Ma+2019}
{Ma}, Y., {Wang}, H., {Zhang}, M., {Li}, C., \& {Yang}, J. 2019, \apj, 878, 44

\bibitem[{{MacLaren} {et~al.}(1988){MacLaren}, {Richardson}, \&
  {Wolfendale}}]{MacLaren+1988}
{MacLaren}, I., {Richardson}, K.~M., \& {Wolfendale}, A.~W. 1988, \apj, 333,
  821

\bibitem[{{Matthews} {et~al.}(1998){Matthews}, {Wallace}, \&
  {Taylor}}]{Matthews+1998}
{Matthews}, B.~C., {Wallace}, B.~J., \& {Taylor}, A.~R. 1998, \apj, 493, 312

\bibitem[{{Maxted} {et~al.}(2018){Maxted}, {Burton}, {Braiding}, {Rowell},
  {Sano}, {Voisin}, {Capasso}, {P{\"u}hlhofer}, \& {Fukui}}]{Maxted+2018}
{Maxted}, N., {Burton}, M., {Braiding}, C., {et~al.} 2018, \mnras, 474, 662

\bibitem[{{McKee} \& {Cowie}(1975)}]{McKeeCowie1975}
{McKee}, C.~F., \& {Cowie}, L.~L. 1975, \apj, 195, 715

\bibitem[{{McKee} \& {Ostriker}(1977)}]{McKeeOstriker1977}
{McKee}, C.~F., \& {Ostriker}, J.~P. 1977, \apj, 218, 148

\bibitem[{Orlando {et~al.}(2005)Orlando, Peres, Reale, Bocchino, Rosner, Plewa,
  \& Siegel}]{Orlando+2005}
Orlando, S., Peres, G., Reale, F., {et~al.} 2005, A{\&}A, 444, 505

\bibitem[{{Pastorello} {et~al.}(2004){Pastorello}, {Zampieri}, {Turatto},
  {Cappellaro}, {Meikle}, {Benetti}, {Branch}, {Baron}, {Patat}, {Armstrong},
  {Altavilla}, {Salvo}, \& {Riello}}]{Pastorello+2004}
{Pastorello}, A., {Zampieri}, L., {Turatto}, M., {et~al.} 2004, \mnras, 347, 74

\bibitem[{{Reich} {et~al.}(1990){Reich}, {Fuerst}, {Reich}, \&
  {Reif}}]{Reich+1990}
{Reich}, W., {Fuerst}, E., {Reich}, P., \& {Reif}, K. 1990, \aaps, 85, 633

\bibitem[{{Reich} {et~al.}(1988){Reich}, {F{\"u}rst}, {Reich}, \&
  {Junkes}}]{Reich+1988}
{Reich}, W., {F{\"u}rst}, E., {Reich}, P., \& {Junkes}, N. 1988, in IAU Colloq.
  101: Supernova Remnants and the Interstellar Medium, ed. R.~S. {Roger} \&
  T.~L. {Landecker}, 293

\bibitem[{{Reid} {et~al.}(2016){Reid}, {Dame}, {Menten}, \&
  {Brunthaler}}]{Reid+2016}
{Reid}, M.~J., {Dame}, T.~M., {Menten}, K.~M., \& {Brunthaler}, A. 2016, \apj,
  823, 77

\bibitem[{{Reid} {et~al.}(2019){Reid}, {Menten}, {Brunthaler}, {Zheng}, {Dame},
  {Xu}, {Li}, {Sakai}, {Wu}, {Immer}, {Zhang}, {Sanna}, {Moscadelli}, {Rygl},
  {Bartkiewicz}, {Hu}, {Quiroga-Nu{\~n}ez}, \& {van Langevelde}}]{Reid+2019}
{Reid}, M.~J., {Menten}, K.~M., {Brunthaler}, A., {et~al.} 2019, \apj, 885, 131

\bibitem[{{Ren} {et~al.}(2014){Ren}, {Li}, \& {Chapman}}]{Ren+2014}
{Ren}, Z., {Li}, D., \& {Chapman}, N. 2014, \apj, 788, 172

\bibitem[{Renaud {et~al.}(2010)Renaud, Marandon, Gotthelf, Rodriguez, Terrier,
  Mattana, Lebrun, Tomsick, \& Manchester}]{Renaud+2010}
Renaud, M., Marandon, V., Gotthelf, E.~V., {et~al.} 2010, ApJ, 716, 663

\bibitem[{Ripple {et~al.}(2013)Ripple, Heyer, Gutermuth, Snell, \&
  Brunt}]{Ripple+2013}
Ripple, F., Heyer, M.~H., Gutermuth, R., Snell, R.~L., \& Brunt, C.~M. 2013,
  \mnras, 431, 1296

\bibitem[{{Shan} {et~al.}(2012){Shan}, {Yang}, {Shi}, {Yao}, {Zuo}, {Lin},
  {Chen}, {Zhang}, {Duan}, {Cao}, {Li}, {Li}, {Liu}, \& {Zhong}}]{Shan+2012}
{Shan}, W.~L., {Yang}, J., {Shi}, S.~C., {et~al.} 2012, IEEE Transactions on
  Terahertz Science and Technology, 2, 593

\bibitem[{{Spitzer}(1978)}]{Spitzer1978}
{Spitzer}, L. 1978, {Physical processes in the interstellar medium} (New York:
  Wiley)

\bibitem[{{Stil} \& {Irwin}(2001)}]{StilIrwin2001}
{Stil}, J.~M., \& {Irwin}, J.~A. 2001, \apj, 563, 816

\bibitem[{{Su} {et~al.}(2014){Su}, {Yang}, {Zhou}, {Zhou}, \& {Chen}}]{Su+2014}
{Su}, Y., {Yang}, J., {Zhou}, X., {Zhou}, P., \& {Chen}, Y. 2014, \apj, 796,
  122

\bibitem[{{Su} {et~al.}(2017{\natexlab{a}}){Su}, {Zhou}, {Yang}, {Chen},
  {Chen}, {Gong}, \& {Zhang}}]{Su+2017b}
{Su}, Y., {Zhou}, X., {Yang}, J., {et~al.} 2017{\natexlab{a}}, \apj, 845, 48

\bibitem[{{Su} {et~al.}(2018){Su}, {Zhou}, {Yang}, {Chen}, {Chen}, \&
  {Zhang}}]{Su+2018}
---. 2018, \apj, 863, 103

\bibitem[{{Su} {et~al.}(2017{\natexlab{b}}){Su}, {Zhou}, {Yang}, {Chen},
  {Chen}, {Liu}, {Wang}, {Li}, \& {Zhang}}]{Su+2017a}
---. 2017{\natexlab{b}}, \apj, 836, 211

\bibitem[{{Sun} {et~al.}(2011){Sun}, {Reich}, {Han}, {Reich}, {Wielebinski},
  {Wang}, \& {M{\"u}ller}}]{Sun+2011}
{Sun}, X.~H., {Reich}, W., {Han}, J.~L., {et~al.} 2011, \aap, 527, A74

\bibitem[{{Tian} {et~al.}(2010){Tian}, {Li}, {Leahy}, {Yang}, {Yang},
  {Yamazaki}, \& {Lu}}]{Tian+2010}
{Tian}, W.~W., {Li}, Z., {Leahy}, D.~A., {et~al.} 2010, \apj, 712, 790

\bibitem[{{Vasisht} {et~al.}(1994){Vasisht}, {Kulkarni}, {Frail}, \&
  {Greiner}}]{Vasisht+1994}
{Vasisht}, G., {Kulkarni}, S.~R., {Frail}, D.~A., \& {Greiner}, J. 1994, \apjl,
  431, L35

\bibitem[{{Voisin} {et~al.}(2016){Voisin}, {Rowell}, {Burton}, {Walsh},
  {Fukui}, \& {Aharonian}}]{Voisin+2016}
{Voisin}, F., {Rowell}, G., {Burton}, M.~G., {et~al.} 2016, \mnras, 458, 2813

\bibitem[{{Xiao} \& {Zhu}(2014)}]{XiaoZhu2014}
{Xiao}, L., \& {Zhu}, M. 2014, \mnras, 438, 1081

\bibitem[{{Yu} {et~al.}(2019){Yu}, {Chen}, {Jiang}, \& {Zijlstra}}]{Yu+2019}
{Yu}, B., {Chen}, B.~Q., {Jiang}, B.~W., \& {Zijlstra}, A. 2019, \mnras, 1915

\bibitem[{{Zhang} {et~al.}(2015){Zhang}, {Chen}, {Su}, {Zhou}, {Pannuti}, \&
  {Zhou}}]{Zhang+2015}
{Zhang}, G.-Y., {Chen}, Y., {Su}, Y., {et~al.} 2015, \apj, 799, 103

\bibitem[{{Zhou} {et~al.}(2016{\natexlab{a}}){Zhou}, {Chen}, {Safi-Harb},
  {Zhou}, {Sun}, {Zhang}, \& {Zhang}}]{Zhoup+2016a}
{Zhou}, P., {Chen}, Y., {Safi-Harb}, S., {et~al.} 2016{\natexlab{a}}, \apj,
  831, 192

\bibitem[{{Zhou} {et~al.}(2016{\natexlab{b}}){Zhou}, {Chen}, {Zhang}, {Li},
  {Safi-Harb}, {Zhou}, \& {Zhang}}]{Zhoup+2016}
{Zhou}, P., {Chen}, Y., {Zhang}, Z.-Y., {et~al.} 2016{\natexlab{b}}, \apj, 826,
  34

\bibitem[{{Zhou} {et~al.}(2014){Zhou}, {Yang}, {Fang}, \& {Su}}]{Zhou+2014}
{Zhou}, X., {Yang}, J., {Fang}, M., \& {Su}, Y. 2014, \apj, 791, 109

\bibitem[{{Zhou} {et~al.}(2016{\natexlab{c}}){Zhou}, {Yang}, {Fang}, {Su},
  {Sun}, \& {Chen}}]{Zhou+2016}
{Zhou}, X., {Yang}, J., {Fang}, M., {et~al.} 2016{\natexlab{c}}, \apj, 833, 4

\bibitem[{{Zhu} {et~al.}(2014){Zhu}, {Tian}, \& {Zuo}}]{Zhu+2014}
{Zhu}, H., {Tian}, W.~W., \& {Zuo}, P. 2014, \apj, 793, 95

\bibitem[{{Zuo} {et~al.}(2011){Zuo}, {Li}, {Sun}, {Yang}, {Li}, {Xu}, {He},
  {Fan}, \& {Fan}}]{Zuo+2011}
{Zuo}, Y.~X., {Li}, Y., {Sun}, J.~X., {et~al.} 2011, Acta Astronomica Sinica,
  52, 152

\end{thebibliography}

\end{document}